# Spatially variant arbitrary polarization shaping for optical skyrmions generation

YU-HAO LEI,[1, 2, 3, 4†] ZI-TING LIU,[1†] JUN-QUAN GUO,[3, 4†] XIN ZHANG,[1] MINXIN YE,[2] QI-DAI CHEN,[1] KE-MI XU,[3, 4*] SHIH-CHI CHEN[2*] AND LEI WANG[1*]

[1]*State Key Laboratory of Integrated Optoelectronics, College of Electronic Science and Engineering, Jilin University, Changchun 130012, China*
[2]*Department of Mechanical and Automation Engineering, The Chinese University of Hong Kong, Shatin, Hong Kong SAR, China*
[3]*MIIT Key Laboratory of Complex-field Intelligent Exploration, School of Optics and Photonics, Beijing Institute of Technology, Beijing 100081, China*
[4]*Beijing Institute of Technology, Zhuhai, China*
†*Yu-Hao Lei, Zi-Qing Liu and Jun-Quan Guo contributed equally to this work.*
[*]*Correspondence: xukemi@bit.edu.cn; scchen@mae.cuhk.hk; leiwang1987@jlu.edu.cn.*

**Abstract:** Polarization is fundamental degree of freedom of light, crucial for applications from imaging to quantum optics. Although numerous methods can spatially manipulate the polarization orientation, e.g., for generating vector beam, achieving simultaneous, spatially resolved control of both the orientation (ψ) and ellipticity (χ) remains challenging. Here, we present an efficient and compact platform to generate spatially variant arbitrary polarization states, e.g., optical skyrmions, in free space, using cascaded spatially variant waveplates in a single silica glass plate via ultrafast laser direct writing. We propose two configurations: (1) a spatially variant half-waveplate (S-HWP) and followed by a quarter-waveplate (S-HWP); and (2) two cascaded spatially variant quarter-waveplates (S-QWPs). Design rules linking the target polarization parameters (orientation ψ and ellipticity χ) to the fast-axis distributions enable spatially resolved arbitrary polarization control. Using these devices, we realize Néel-, Bloch-, and anti-skyrmions of different orders (e.g., second and fourth) and n-π textures (2π and 3π), with polarization-resolved measurements in excellent agreement with simulations. We further demonstrate 3×3 arrays comprising either identical or hybrid skyrmions. This approach enables the realization of spatially variant arbitrary polarization state and scalable skyrmion lattices.

## Introduction

Polarization is a fundamental degree of freedom of light, and the ability to shape it on demand in space underpins advances in imaging, light–matter interactions, optical communications, and quantum information [1-6]. Existing platforms can tailor the polarization orientation to create vector beams, e.g., radially and azimuthally polarized beams, yet simultaneous and spatially resolved control of both orientation (ψ) and ellipticity (χ) remains challenging. Such full-Poincaré control is not merely a technical pursuit but a prerequisite for generating complex topological light fields, most notably optical skyrmions.

Skyrmions [7] were first introduced as topologically stable quasiparticles, which have been predicted and observed in many contexts, such as nucleons [7,8], liquid crystals [9] and magnetic materials [10,11]. Among them, the most well-known type is magnetic skyrmions formed by a magnetization texture [10], paving new avenues in high density storage and low-energy magnetic memory [12,13]. In optics, skyrmions manifest as sophisticated polarization textures characterized by an integer topological invariant (skyrmion number), which provides inherent robustness against perturbations. More recently, optical skyrmions [14] have been observed in specially shaped surface plasmon polariton (SPP) [15] and spin distribution of an SPP field [16]. Given by the feature of ultrasmall and topological stability [14], optical skyrmions are promising for applications in light-matter interactions, super-resolution imaging, optical communications and quantum information [17].

Motivated by the scientific and technological promise of optical skyrmions, researchers have pursued numerous strategies for their free-space generation [18-21]. However, practical implementations are often constrained by the sophisticated and bulky spatial light modulation systems. Recently, compact unitary spin-orbit waveplates in silica glass, featuring ultrafast laser written spatially varying retardance and slow-axis orientation, have been used to generate optical skyrmions [22,23]; however, prescribing an arbitrary retardance distribution is technically challenging and can introduce fabrication artifacts [24]). In general, generating optical skyrmions requires precise, spatially resolved control of both polarization orientation (ψ) and ellipticity (χ), while simultaneous modulation of both polarization orientation and ellipticity across the space field remains challenging with existing platforms.

Here, we propose and demonstrate a compact platform to generate spatially variant arbitrary polarization, e.g., optical skyrmions, in free space using two cascaded spatially variant waveplates (S-WPs) written by ultrafast laser in a single silica glass substrate. Starting from the target skyrmion texture, specified by the spatial maps of polarization orientation (ψ) and ellipticity ($\chi$), we calculate the position-resolved fast-axis distributions of two waveplates via the Stokes vector formalism and Müller matrices analysis, and then realize them by ultrafast laser writing of nanopores based birefringent modifications [25]. The fabricated devices generate skyrmions of different types (Néel-, Bloch-, and anti-), topological charges (up to $N_{sk}$ = 4) and n-$\pi$ textures; the measured polarization-component images ($I_0$, $I_{45}$, $I_{90}$, $I_{135}$, $I_L$, $I_R$) are in good agreement with simulations. Finally, we further demonstrate 3×3 arrays of optical skyrmion in both uniform and hybrid configurations. We anticipate that high intensity femtosecond skyrmion beam can be generated by our device for direct writing of solid materials.

## Results and discussion

### Design of cascaded spatial-variant waveplate

The optical skyrmions, i.e., Stokes skyrmions, with spatial-variant polarization states covers the whole Poincaré sphere of polarization, which is expressed by the polarization orientation angle (0° ≤ ψ < 180°) and ellipticity angle (-45° ≤ χ ≤ 45°) [Fig. 1a]. The normalized Stokes vector field can be defined as

$$S = [1, \sin(\theta_r)\cos(N_{sk}\varphi)\,,\ \sin(\theta_r)\sin(N_{sk}\varphi)\,, \cos(\theta_r)]^T, \tag{1}$$

where $\theta_r$ is the polar angle (0°-180°) on the Poincaré sphere (controlling the polarization ellipticity χ), $N_{sk}$ is the charge of skyrmions and $\varphi$ is the azimuth angle in the transverse plane (controlling the polarization orientation ψ). Note that the polarization orientation (ψ) and ellipticity (χ) can be calculated through Stokes parameters:

$$\psi = \frac{1}{2}\arctan\left(\frac{S_2}{S_1}\right),\ \ \chi = \frac{1}{2}\arcsin(S_3)\,. \tag{2}$$

Therefore, the generation of optical skyrmions is equivalent to producing arbitrary polarization states at different spatial positions. To realize any polarization state characterized by an orientation angle ψ (0° ≤ ψ < 180°) and an ellipticity angle χ (-45° ≤ χ ≤ 45°), we propose two approaches: (1) a combination of a half-wave plate (HWP) and a quarter-wave plate (QWP) with horizontally polarized incident light; and (2) a pair of QWPs with right-hand circularly polarized incident light. The mathematical relationship between the polarization parameters, i.e., orientation ($\psi$) and ellipticity ($\chi$) angles, and the fast-axis angle of wave plates is derived using the Stokes vector formalism and Müller matrices analysis (Supplement 1, section 1).

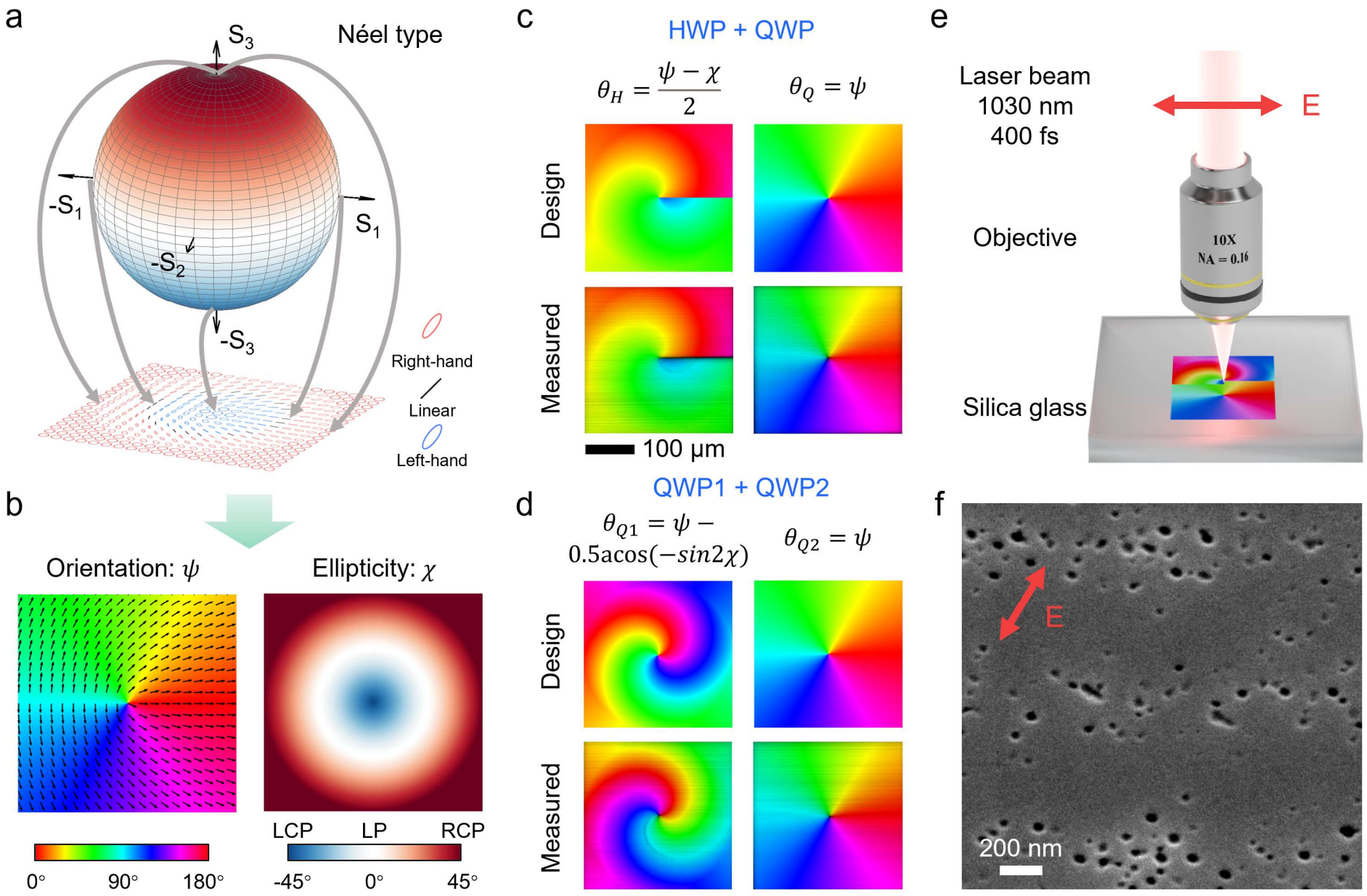

**Fig. 1.** Generation of optical skyrmions by cascaded spatially variant wave plates. (a), Mapping of an optical Stokes skyrmions. (b) Spatial distribution of polarization orientation ($\psi$) and ellipticity ($\chi$) corresponding to the optical skyrmions. (c) Fast-axis distribution of a spatial variant half-wave plate (S-HWP) followed by a quarter-wave plate (S-QWP), with horizontally polarized incident light. (d) A spatial variant S-QWP1 and S-QWP2, with right-hand circularly polarized incident light. In both configurations (c and d), the designed (top row) and experimentally measured (bottom row) fast-axis distributions are shown. The size of S-waveplate was 300 μm by 300 μm. Scale bar: 100 μm. (e) Ultrafast laser writing of cascaded S-QWPs in silica glass. (f) SEM image of the ultrafast-laser-induced nanopores. The doubled arrow shows the azimuth of polarization of the writing laser beam. Scale bar: 200 nm.

For a configuration consisting of a half-wave plate (HWP) followed by a quarter-wave plate (QWP), the required fast axis orientations to generate an arbitrary output polarization state are given by

$$\theta_H = \frac{\psi-\chi}{2}, \theta_Q = -\chi \tag{3}$$

as shown in Fig. 1c. (Alternatively, if the light passes through an QWP followed by a HWP, the required fast axis orientations are $\theta_Q = \psi$ and $\theta_H = \frac{\psi-\chi}{2}$).

However, fabricating a spatially variant HWP requires approximately twice the processing time compared to that of a QWP, and a clear laser writing artifact, i.e., a dark horizontal line, is observed in the spatial-variant HWP [Fig. 1c]. We therefore further derive a second, more fabrication-friendly solution based on two QWPs with right-hand circularly polarized incident light. In this case, the target polarization state can also be generated with:

$$\theta_{Q1} = \psi - 0.5 \times \arccos(-\sin 2\chi), \theta_{Q2} = \psi, \tag{4}$$

as detailed in Supplement 1, Section 1. (Note that arbitrary polarization state can also be generated using two QWPs with linearly polarized incident light; however, the required fast-axis orientations are too complicated to be presented by an analytical equation.). In addition, the strategy of using quarter-wave plates not only reduce processing time, but also avoids obvious writing artifact of dark line [Fig. 1d]. The S-QWPs were fabricated in silica glass using ultrafast laser writing [Fig. 1e]; details of the writing setup are provided in our previous work [25,26]. To reveal the origin of the laser-induced birefringence, the written regions were polished and etched, followed by SEM characterization [Fig. 1f]. The SEM image shows randomly distributed anisotropic nanopores with a width of about 40-50 nm, while these nanopores were elongated perpendicular to the polarization direction of the writing laser beam [25], enabling the construction of S-waveplates in silica glass.

**Generation of optical skyrmions via cascaded S-QWPs**

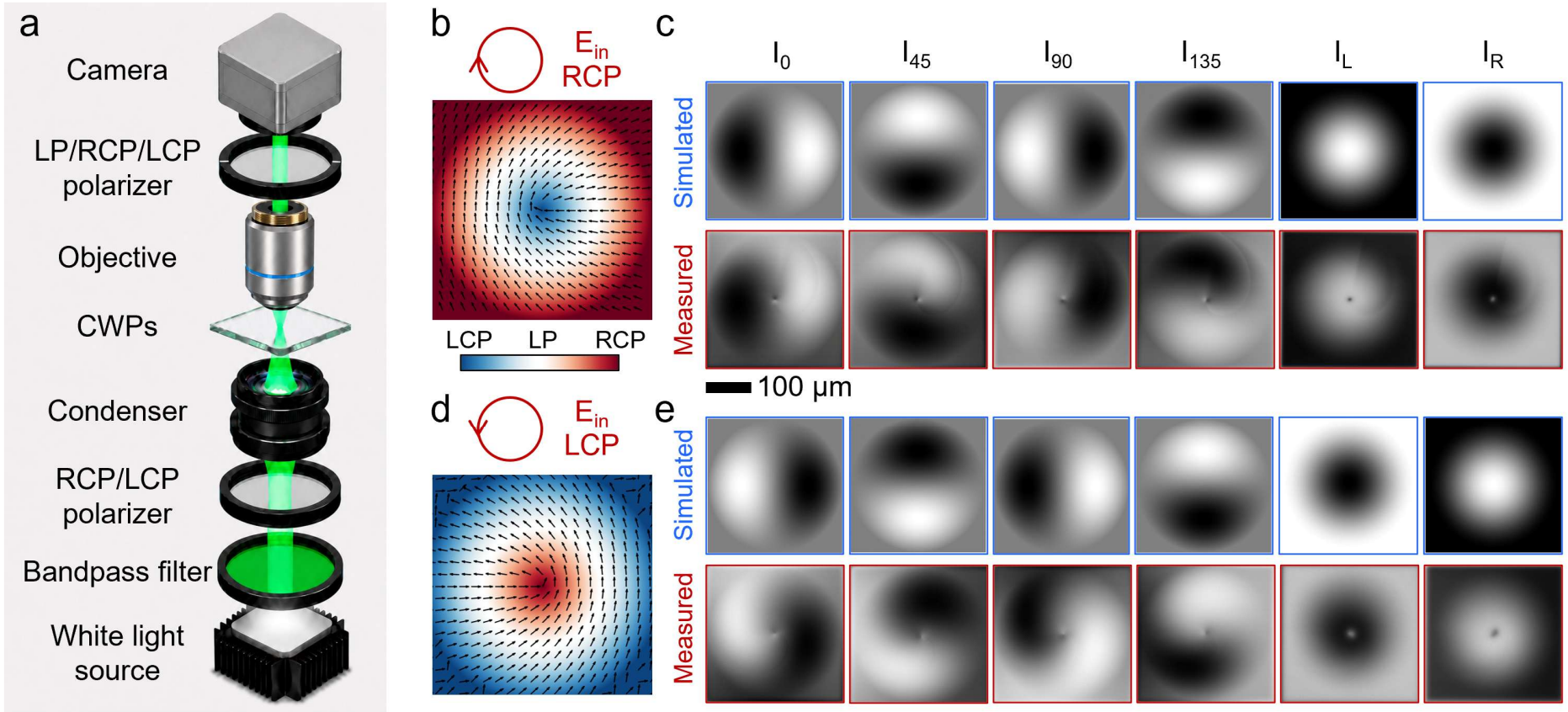

**Fig. 2.** Néel-type optical skyrmions generated with different incident polarization states. (a) Experimental setup for polarization-resolved characterization of the optical skyrmions generated by the fabricated cascaded wave plates (CWPs). A white-light source, a bandpass filter, and a circular polarizer are used to prepare the incident polarization state, which is then focused onto the CWPs by a condenser. The transmitted field is collected by an objective, analyzed by a linear or circular polarization analyzer, and captured by a camera. (b) Spatial distribution of polarization orientation ($\psi$) and ellipticity ($\chi$) for a Néel-type optical skyrmions generated with RCP incident light. (c) Simulated (top) and measured (bottom) intensity distribution of six polarization components ($I_0$, $I_{45}$, $I_{90}$, $I_{135}$, $I_L$, $I_R$) of the optical skyrmions in (b). (d) Spatial polarization distribution of a Néel type Skyrmion generated with LCP incident light. (e) Corresponding simulated and measured polarization components of the skyrmions in d. Scale bar: 100 µm.

To experimentally demonstrate the generation of optical skyrmions, we fabricated a pair of cascaded spatially variant quarter-wave plates (S-QWPs; 300-µm diameter each) according to the designs in Fig. 1d and then analyzed the output under different incident polarization states [Fig. 2a]. With right-hand circularly polarized (RCP) incident light, the device, cascaded S-QWPs, produces a Néel-type optical skyrmion ($N_{sk} = +1$) with the RCP component localized at the centre of the beam [Fig. 2b]. Moreover, intensity images of six polarization components ($I_0$, $I_{45}$, $I_{90}$, $I_{135}$, $I_L$, $I_R$) show excellent agreement between simulation (top row) and measurement (bottom row), confirming the designed vector field [Fig. 2c].

On the other hand, reversing the incident helicity to left-hand circular polarization (LCP) flips the skyrmion number to $N_{sk} = -1$, while preserving the characteristic Néel-type texture [Fig. 2d]. The corresponding simulated and measured intensity image of different polarization components again match closely [Fig. 2e], validating the spatially variant polarization field. This helicity-dependent switchability of the generated optical skyrmions is similar to that report for Q-plate, where switching is controlled by the polarization orientation of the incident linearly polarized (LP) light [22]. Beyond optical skyrmions, the same device illuminated with linearly polarized light (e.g., horizontally polarized) generates a "Tai-Chi"-like vector beam featuring a spiral polarization pattern (Fig. S1, Supplement 1 section 2). And its measured polarization-component images agree well with simulations. Such structured polarization fields may enhance the efficiency of laser processing and improve the performance of multiplexed optical data storage.

**Different orders and types of optical skyrmions**

We also demonstrated the generation of optical skyrmions across multiple orders and types, highlighting the broad applicability of our method. For example, higher-order optical skyrmions are polarization textures with an integer-valued skyrmion number $| N_{sk} | > 1$. Specifically, we realized the second-order and forth-order Néel-type skyrmions [Fig. 3a, c] by changing the fast-axis distribution of cascaded S-QWPs (Fig. S2, Supplement 1 section 3). Their measured (bottom row) polarization-component images agree well with simulations (top row), confirming the accuracy of our design and fabrication [Fig. 3b, d]. Apart from high-order skyrmions, we also showed the n-π skyrmions, whose radial polarization twist from the core to the periphery accumulates to nπ.

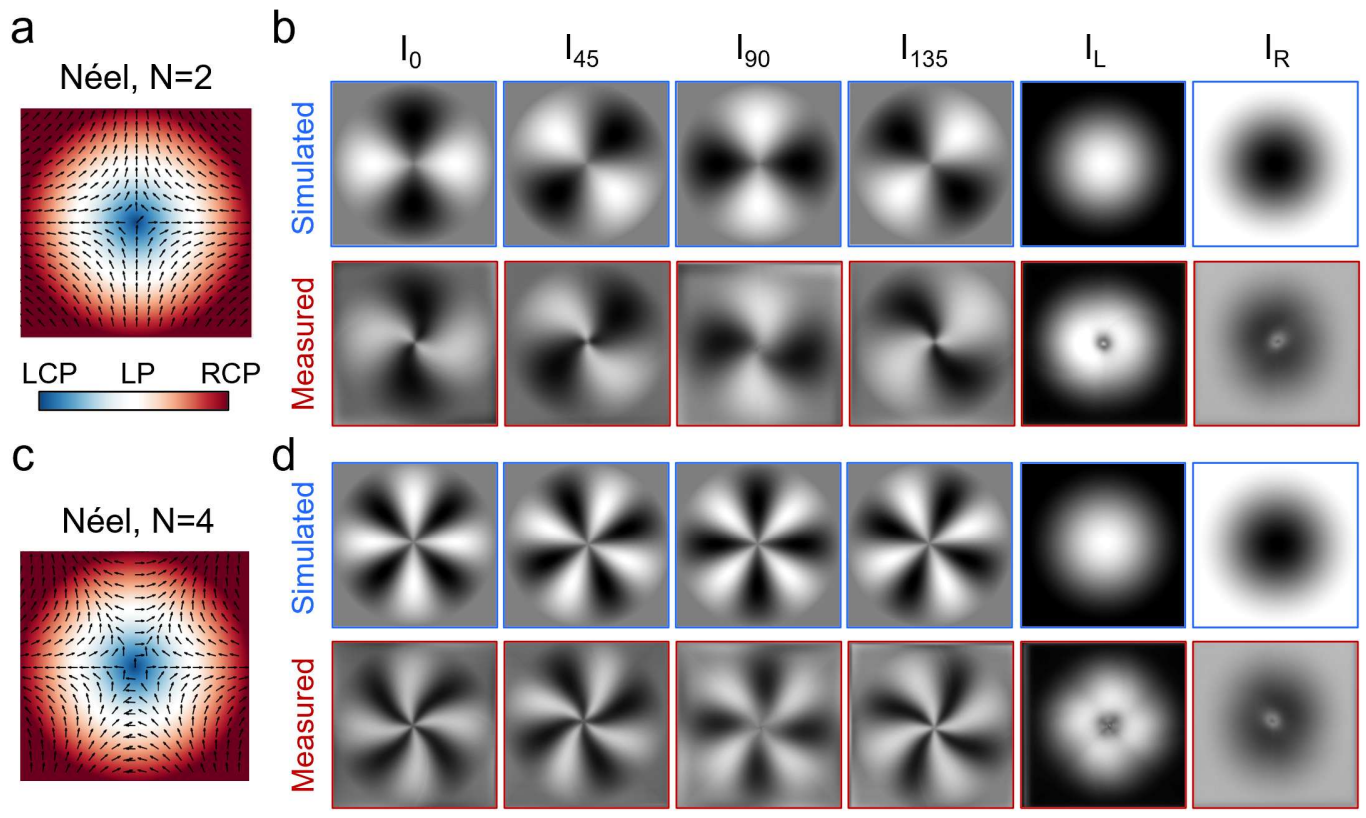

**Fig. 3.** High-order Néel-type optical skyrmions. (a) Spatial polarization distribution of a second-order Néel-type Skyrmion with topological number N = 2. (b) Simulated (top) and measured (bottom) intensity distribution of six polarization components ($I_0$, $I_{45}$, $I_{90}$, $I_{135}$, $I_L$, $I_R$) of the optical skyrmions in (a). (c) Spatial polarization distribution of a fourth-order Néel-type skyrmion (N = 4). (d) Corresponding simulated and measured polarization components of the Skyrmions in (c).

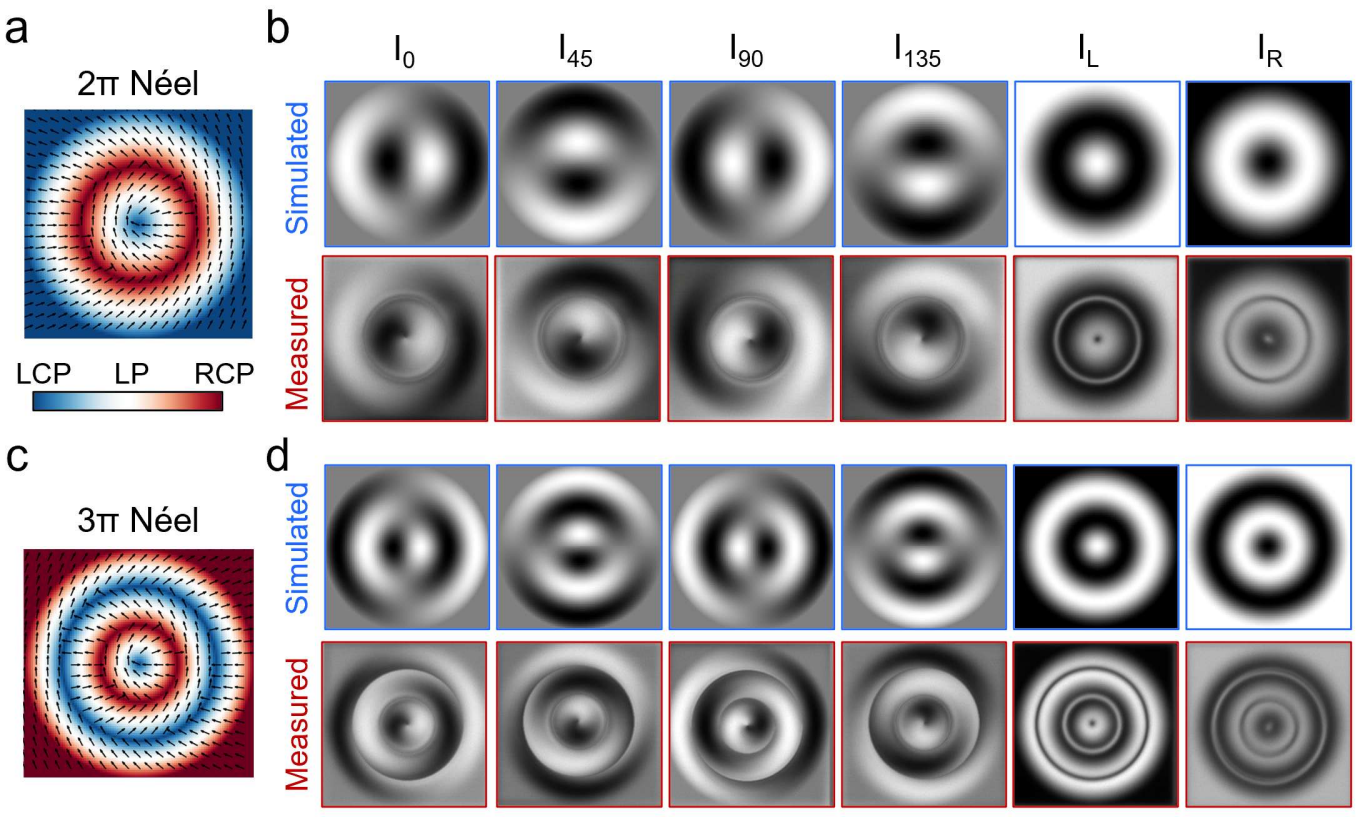

**Fig. 4.** 2π and 3π Néel-type optical skyrmions. (a) Spatial polarization distribution of a 2π Néel-type skyrmion. (b) Simulated (top) and measured (bottom) intensity distribution of six polarization components ($I_0$, $I_{45}$, $I_{90}$, $I_{135}$, $I_L$, $I_R$) of the optical skyrmions in (a). (c) Spatial polarization distribution of a 3π Néel-type skyrmion. (d) Corresponding simulated and measured polarization components of the skyrmions in (c).

Using the analytical formulation for 2π and 3π Néel-type skyrmions [Fig. 4a, c], we derived the corresponding fast-axis distributions for the cascaded S-QWPs and fabricated them experimentally in silica glass (Fig. S3, Supplement 1 section 3). The measured polarization-component images closely match the simulations [Fig. 4b, d], validating the target polarization texture of n-π skyrmions.

Beyond these Néel-type skyrmions, we further demonstrate the generation of Bloth-type and anti-skyrmions [14]. The normalized Stokes vector field of skyrmion can be defined as $S = [1, \sin(\theta_r)\cos(N_{sk}\varphi + \varphi_1),\ \sin(\theta_r)\sin(N_{sk}\varphi + \varphi_2), \cos(\theta_r)]^T$, where $\varphi_1$ and $\varphi_2$ are phase offsets. In the case of $\varphi_1 = \varphi_2 = 0$, the skyrmions are classified as Néel-type and have a 'hedgehog' texture around the quasiparticle. For $\varphi_1 = \varphi_2 = \pi/2$, the skyrmions are classified as Bloth-type with a vortex texture around the center. For $\varphi_1 = 0, \varphi_2 = \pi$, the skyrmions are defined as anti-skyrmions with a saddle texture.

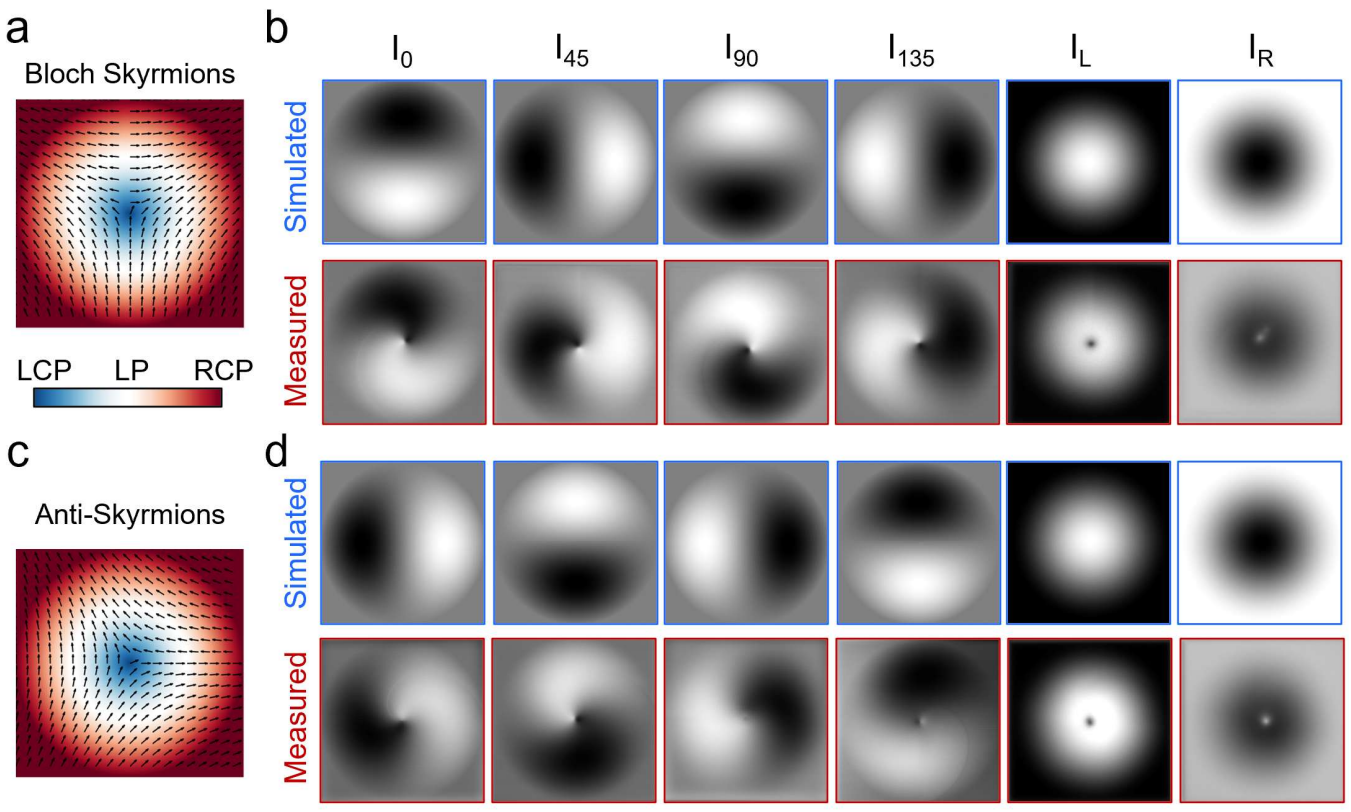

**Fig. 5.** Bloch-type and anti-skyrmions. (a) Spatial polarization distribution of a Bloch-type skyrmion. (b) Simulated (top) and measured (bottom) intensity distribution of six polarization components ($I_0$, $I_{45}$, $I_{90}$, $I_{135}$, $I_L$, $I_R$) of the optical skyrmions in (a). (c) Spatial polarization distribution of anti-Skyrmion. (d) Corresponding simulated and measured polarization components of the skyrmions in (c).

Using these closed-form expressions, we derive the spatial distribution of the polarization orientation and ellipticity angles for Bloch-type and anti-skyrmions [Fig. 5a, c], from which the fast-axis distributions for the corresponding S-QWPs are calculated (Fig. S4, Supplement 1 section 3). We experimentally generated both Bloch-type and anti-skyrmions and their measured polarization-component images agree well with simulations [Fig. 5b, d], confirming the successful generation of different optical skyrmions types.

**Arrays of optical skyrmions**

We also experimentally demonstrate arrays of optical skyrmions ("skyrmion crystals"), enabled by the high customizability of our cascaded S-QWPs. Specifically, we first designed two 3×3 configurations: (1) an array of identical Néel-type skyrmions [Fig. 6a]; and (2) a hybrid array combing Néel- and Bloch-type skyrmions with different skyrmion numbers, $N_{sk}$=1, 2, and 3 [Fig. 6c]. Subsequently, the corresponding cascaded S-QWPs were fabricated in silica glass with an overall footprint of 900 µm × 900 µm, with each skyrmion unit measuring 300 µm × 300 µm. The measured the polarization-resolved intensity distribution of the skyrmions arrays closely match simulations [Fig. 6b, d], validating the realization of both the uniform and hybrid skyrmions array. Moreover, as the skyrmion textures are encoded in the generated optical field, their lateral dimensions can in principle be further reduced through optical demagnification or tight focusing, offering a route to more compact skyrmion lattices.

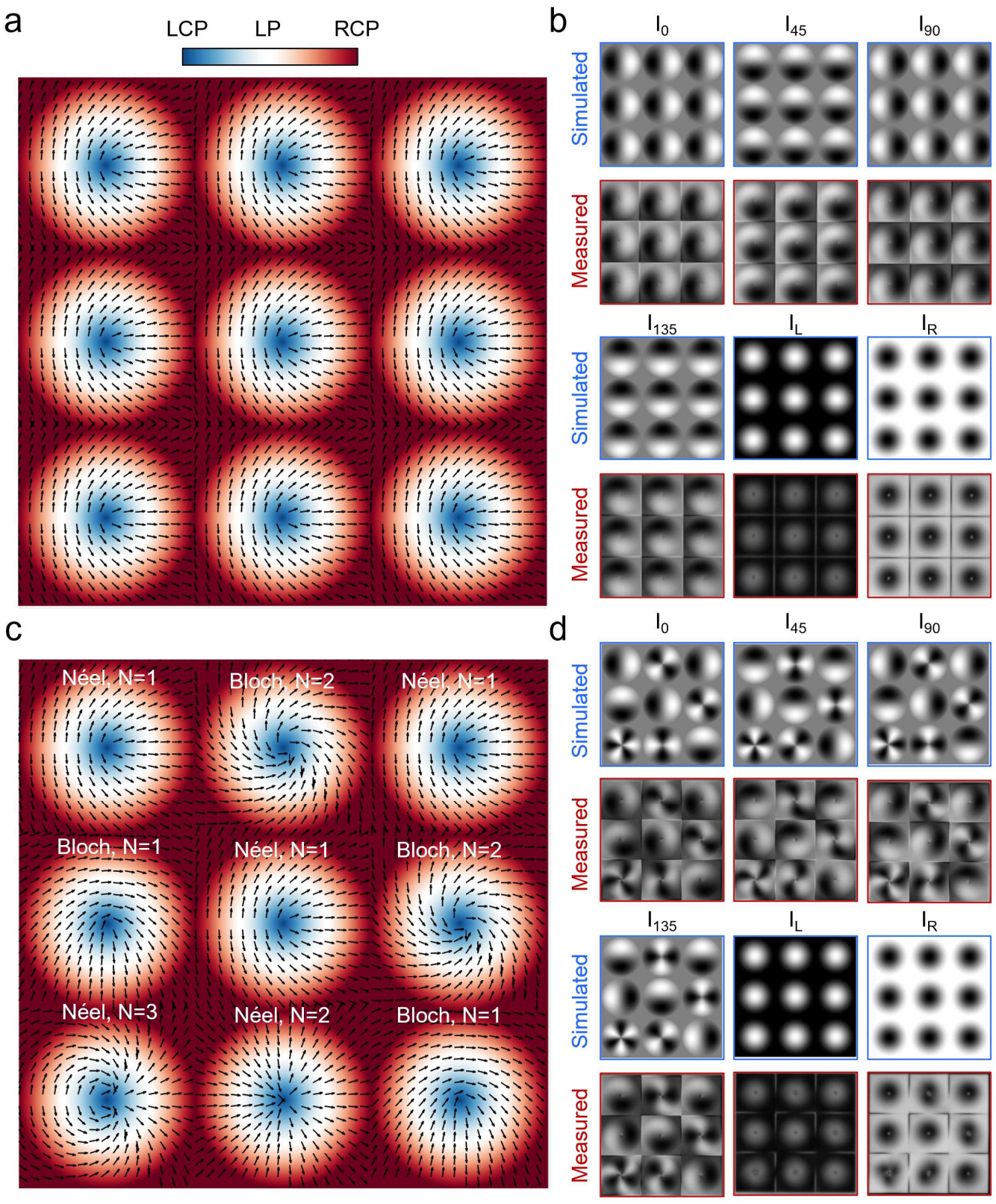

**Fig. 6.** Arrays of optical skyrmions. (a) Simulated polarization maps for a 3×3 array of identical skyrmions. The background color encodes ellipticity and handedness; black dashes indicate the local polarization azimuth. (b) Simulated and measured intensity distributions of six polarization components ($I_0$, $I_{45}$, $I_{90}$, $I_{135}$, $I_L$, $I_R$) for the array in (a). (c) Simulated polarization map for a 3×3 hybrid array including Néel- and Bloch-type skyrmions with topological numbers $N_{sk} = 1, 2$, and 3. (d) Simulated and measured polarization resolved intensity images for the array in (c).

## Conclusion

In conclusion, we demonstrate a general route to spatially resolved arbitrary polarization fields, taking optical skyrmions as showcases, using two cascaded spatially variant waveplates fabricated via ultrafast laser direct writing. The required fast-axis maps of S-QWPs are analytically retrieved from the target polarization orientation ($\psi$) and ellipticity ($\chi$), and then realized by writing of birefringent modifications in silica glass. These devices were used to generate different types and orders of skyrmions as well as 3×3 arrays; polarization resolved measurements agree well with simulations. More broadly, this work goes beyond skyrmion generation itself: to the best of our knowledge, it is the first to formulate free-space optical skyrmion generation as an analytical arbitrary polarization shaping problem solved with cascaded spatially variant wave plates, which is readily to be used in liquid-crystal based devices and other metasurfaces.

In future work, the topological degree of freedom in optical skyrmions may serve as an additional channel for multidimensional optical data storage [6,27], thereby enabling higher-capacity encoding. Moreover, the cascaded S-QWPs device with high optical damage threshold is compatible with high power ultraviolet (UV) laser beams [28], supporting the generation of UV beam with spatially random polarization distribution, which is desired for laser-driven inertial confinement fusion applications [29]. In addition, we anticipate that the device can be used to produce high-power infrared femtosecond skyrmion beams and then could be converted to extreme ultraviolet attosecond skyrmion fields via high-order harmonic generation [30].

## Methods

### Sample fabrication

The light source of the ultrafast laser writing setup was a Yb:KGW femtosecond laser system (Carbide, Light Conversion), operating at a wavelength of 1030 nm, with a repetition rate of 200 kHz and a pulse duration of 400 fs. The polarization of the writing laser was controlled by the combination of an electro-optic modulator and a quarter-wave plate. The laser beam was focused into a silica glass sample at a depth of 170 μm using an aspheric lens with a numerical aperture (NA) of 0.16, and the sample was mounted on a three-axis air-bearing translation stage. The sample was translated at a speed of 6 mm/s, resulting in a pulse density of about 33 pulses per micrometer. The spatially variant waveplates were written in a raster scanning manner with a line internal of 1 μm. The pulse energy was 0.95 μJ and the pulse duration was 400 fs, yielding a fluence of 4.1 J/cm$^2$ and an intensity of 10.3 TW/cm$^2$. Notably, a single birefringent-modification layer provides a retardance of approximately 65 nm, so two such layers were written to form a spatially variant quarter-wave plate (~130 nm) for light at a wavelength of 532 nm.

### Sample characterization

To characterize and record the birefringence, including retardance and slow axis azimuth, of the laser-written modifications, a commercial birefringence measurement system (Oosight, Hamilton Thorne) that operates at 546 nm with an integrated Olympus BX53 optical microscope was used. The same optical microscope was used to characterize the generated optical skyrmions, and different polarizers, i.e., linear and circular polarizers, were employed to capture intensity images of six polarization components. Note that a band pass filter with a central wavelength of 532 nm (MBF10-532-10, Lebtk) and a right-hand circular polarizer (RCP25-VIS-M) were inserted after the white light source to generate the required RCP incident light. In addition, after polishing and etching with a KOH solution (1 mol/L) for 24 hours, the laser-written nanopores were characterized by a scanning electron microscope (SEM, JEOL JSM-7900F).

### Supplementary Information

The online version contains supplementary material available at

### Authors' contributions

Y.-H. L. and L.W. conceived the study. Y.-H. L. and Z.-T. L. setup the optical system and performed the laser writing experiments. Y.-H. L. and J.-Q. G. simulated and measured the generated Skyrmions. X. Z. carried out the SEM characterization. M. Y. and Q.-D. C. assisted in designing the experiments. K.-M. X., S.-C. C. and L. W. supervised the project. Y.-H. L. prepared the original manuscript with revisions from all authors.

### Funding

This work was supported by National Key Research and Development Program of China (2024YFB4505100), National Natural Science Foundation of China (62175086), Fundamental and Interdisciplinary Disciplines Breakthrough Plan of the Ministry of Education of China, Research Grants Council (Hong Kong) (14212025), InnoHK Centre projects funded by the Innovation and Technology Commission (ITC RC/IHK/1/3).

### Data availability

The datasets used and/or analyzed during the current study are available from the corresponding author on reasonable request.

### Declarations

### Competing interests

The authors declare no conflicts of interest.